\documentclass[notitlepage, 
bibnotes,
prb,10pt, 
twocolumn,
superscriptaddress,
amsmath,amssymb,floatfix, aps]{revtex4-2}

\usepackage{amssymb}
\usepackage{amsfonts}
\usepackage{mathrsfs}
\usepackage{empheq}
\usepackage{tikz}
\usepackage{tikz-cd}
\usepackage{amsmath}
\usepackage{graphicx}
\usepackage{dsfont}
\usepackage{dcolumn}
\usepackage{bm}
\usepackage{color}
\usepackage{xcolor}
\usepackage{url}
\usepackage{tabularx}
\usepackage{array}
\usepackage{booktabs}
\usepackage{comment}
\usepackage{footnote}
\usepackage{lipsum}
\usepackage{natbib}
\usepackage{xr}
\usepackage[colorlinks,allcolors=blue]{hyperref}
\usepackage[normalem]{ulem}
\usepackage{esint}
\usepackage{multirow}

\UseRawInputEncoding

\makeatletter
\newcommand*{\addFileDependency}[1]{
  \typeout{(#1)}
  \@addtofilelist{#1}
  \IfFileExists{#1}{}{\typeout{No file #1.}}
}
\makeatother

\begin{document}

\title{Magnetic hopfions at room temperature}

\author{Kaixin Zhu}
\affiliation{Beijing National Laboratory for Condensed Matter Physics, Institute of Physics, Chinese Academy of Sciences, Beijing 100190, China}
\affiliation{School of Physical Sciences, University of Chinese Academy of Sciences, Beijing 100049, China}

\author{Wenli Gao}
\affiliation{Beijing National Laboratory for Condensed Matter Physics, Institute of Physics, Chinese Academy of Sciences, Beijing 100190, China}
\affiliation{School of Physics, Northwest University, Xi’an 710069, China}

\author{Zhan Wang}
\affiliation{Beijing National Laboratory for Condensed Matter Physics, Institute of Physics, Chinese Academy of Sciences, Beijing 100190, China}
\affiliation{School of Physical Sciences, University of Chinese Academy of Sciences, Beijing 100049, China}

\author{Shuaishuai Sun}
\affiliation{Beijing National Laboratory for Condensed Matter Physics, Institute of Physics, Chinese Academy of Sciences, Beijing 100190, China}
\affiliation{Songshan Lake Materials Laboratory, Dongguan, Guangdong, 523808, China}

\author{Siyuan Huang}
\affiliation{Beijing National Laboratory for Condensed Matter Physics, Institute of Physics, Chinese Academy of Sciences, Beijing 100190, China}
\affiliation{School of Physical Sciences, University of Chinese Academy of Sciences, Beijing 100049, China}

\author{Junxi Tong}
\affiliation{Beijing National Laboratory for Condensed Matter Physics, Institute of Physics, Chinese Academy of Sciences, Beijing 100190, China}
\affiliation{School of Physical Sciences, University of Chinese Academy of Sciences, Beijing 100049, China}

\author{Jun Li}
\affiliation{Beijing National Laboratory for Condensed Matter Physics, Institute of Physics, Chinese Academy of Sciences, Beijing 100190, China}

\author{Huanfang Tian}
\affiliation{Beijing National Laboratory for Condensed Matter Physics, Institute of Physics, Chinese Academy of Sciences, Beijing 100190, China}

\author{Zian Li}
\thanks{Corresponding author: zianli@gxu.edu.cn}
\affiliation{School of Physical Science and Technology, Guangxi University, Nanning 530004, China}

\author{Huaixin Yang}
\affiliation{Beijing National Laboratory for Condensed Matter Physics, Institute of Physics, Chinese Academy of Sciences, Beijing 100190, China}
\affiliation{School of Physical Sciences, University of Chinese Academy of Sciences, Beijing 100049, China}

\author{Ying Zhang}
\affiliation{Beijing National Laboratory for Condensed Matter Physics, Institute of Physics, Chinese Academy of Sciences, Beijing 100190, China}
\affiliation{School of Physical Sciences, University of Chinese Academy of Sciences, Beijing 100049, China}
\affiliation{Songshan Lake Materials Laboratory, Dongguan, Guangdong, 523808, China}

\author{Olle Eriksson}
\affiliation{Department of Physics and Astronomy, Uppsala University, Uppsala, Sweden}

\author{Filipp N. Rybakov}
\thanks{Corresponding author: philipp.rybakov@physics.uu.se}
\affiliation{Department of Physics and Astronomy, Uppsala University, Uppsala, Sweden}

\author{Nikolai S. Kiselev}
\thanks{Corresponding author: n.kiselev@fz-juelich.de}
\affiliation{Peter Gr\"{u}nberg Institute, Forschungszentrum J\"{u}lich, J\"{u}lich, Germany}

\author{Jianqi Li}
\thanks{Corresponding author: ljq@iphy.ac.cn}
\affiliation{Beijing National Laboratory for Condensed Matter Physics, Institute of Physics, Chinese Academy of Sciences, Beijing 100190, China}
\affiliation{School of Physical Sciences, University of Chinese Academy of Sciences, Beijing 100049, China}
\affiliation{Songshan Lake Materials Laboratory, Dongguan, Guangdong, 523808, China}

\maketitle

\textbf{
Hopfions are three-dimensional (3D) topological solitons predicted to exist in diverse magnetic systems, yet their practical utility has been largely restricted to cryogenic environments. 
Here, we overcome this temperature constraint by demonstrating stable magnetic hopfions in the chiral magnet Co$_8$Zn$_8$Mn$_4$ at and above room temperature.
Using a transmission electron microscope equipped for in situ optical excitation, we generate magnetic hopfions with femtosecond laser pulses. 
Long-term observations further reveal Brownian-like motion at room temperature and thermally activated collapse upon approaching the high-temperature regime.
Together with micromagnetic simulations and homotopy group analysis, our experimental observations uncover the hopfion formation mechanism through the fusion of bimeron pairs.
These findings establish room-temperature magnetic hopfions and provide a framework for their further studies under technologically relevant conditions.
}

Particles can be understood as localized excitations of underlying fields.
In modern quantum field theory, for instance, elementary particles are described as quantized excitations of fundamental fields.
Topological solitons provide a fundamentally different realization of this concept.
Unlike elementary particles, they are nonlinear field configurations whose quantum numbers are topological invariants.
Such objects occur across many branches of physics, from high-energy physics and cosmology to solid-state physics.
A prominent example from high-energy physics is the Skyrme model~\cite{Manton_Sutcliffe_2004}, in which baryons are described as topological solitons of meson fields.
In solid-state physics, representative examples include magnetic vortices, merons, and skyrmions.
Among these, skyrmions have become the archetypal topological particles owing to their experimental realization in a wide range of magnetic materials and their potential for spintronic applications.
In bulk chiral magnets, merons and skyrmions form filamentary textures extending through the sample thickness, commonly referred to as strings or tubes~\cite{rybakov2013, Zhengbraids, Tangskyrmionbundles, Zhu2026bimeron}.
Magnetic vortices, merons, and skyrmions have now been observed in a wide variety of magnetic materials, with several systems supporting these topological textures even at room temperature.

3D topological magnetic solitons, known as hopfions, are often interpreted as closed twisted skyrmion/bimeron strings that may also become linked with one another.
Their 3D nature and more complex morphology makes hopfion nucleation considerably more challenging~\cite{ZhengHopfion}.
Recent studies have shown that ultrashort electric current pulses and femtosecond laser pulses provide efficient routes for nucleating hopfions in chiral magnets of the B20 family~\cite{Li2026hopfions, Chen2026hopfions}.
However, the relatively low Curie temperatures of B20-type chiral magnets have so far restricted experimental observations of hopfions to cryogenic temperatures.
Experimental realizations of magnetic hopfions under ambient conditions therefore remain elusive.

\begin{figure*}[!ht]
\centering
\includegraphics[width=0.85\linewidth]{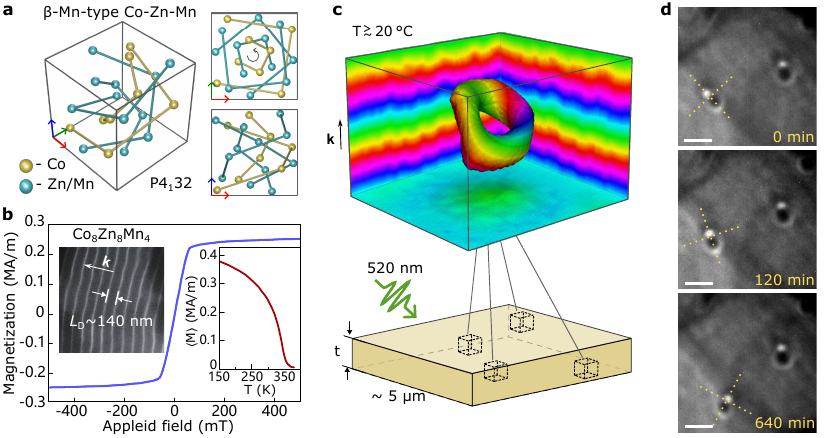}
    \caption{\textbf{Room-temperature magnetic hopfions in Co$_8$Zn$_8$Mn$_4$.}
    \textbf{a}, Crystal structure of the right-handed $\beta$-Mn-type Co--Zn--Mn family. Insets show top and side views of the unit cell.
    \textbf{b}, Room-temperature magnetization curve of a Co$_8$Zn$_8$Mn$_4$ bulk crystal.
    The left inset shows an over-focused Lorentz TEM image of the helical state with modulation period $L_{\rm D}\approx140$~nm.
    The right inset shows the temperature dependence of the magnetization measured after zero-field cooling (ZFC) in an applied field of $100$~mT.
    \textbf{c}, Schematic illustration of sample irradiated by femtosecond laser pulses (bottom) and a room-temperature magnetic hopfion (top). The hopfion is visualized by an isosurface at $m_z=-0.5$ and can reside at various depths within the sample, while the surrounding magnetization is shown along the boundaries of the simulation box. Both include thermal distortions representative of temperature fluctuations.
    \textbf{d}, Under-focus Lorentz TEM images of magnetic hopfions recorded in 190-nm-thick sample at room temperature ($\sim20$~$^\circ$C) at different times.
    The dotted lines highlight the translational and rotational motion of the hopfions.
    Scale bars in all Lorentz TEM images are 200~nm.
    }
    \label{Fig_1}
\end{figure*}

Here, using in situ femtosecond laser excitation and Lorentz transmission electron microscopy (TEM), we demonstrate stable magnetic hopfions in Co$_8$Zn$_8$Mn$_4$ at room temperature.
We directly visualize both the formation and the thermally activated dynamics of magnetic hopfions.
This includes the transformation of a bound pair of oppositely charged bimerons into a 3D hopfion, together with Brownian-like motion.
Their exceptional thermal stability allows hours-long observation at room temperature, while thermally activated collapse becomes observable upon approaching the Curie temperature.
To unambiguously interpret TEM images, we modeled magnetic textures and corresponding theoretical TEM images using a quantitative micromagnetic model.
The resulting digital twins provide access to the full 3D magnetization field and serve as the basis for a rigorous topological analysis.
Using homotopy-group methods, we explain the mechanism of hopfion formation and establish the relationships between hopfions and other textures observed in our experiment, including (bi)merons and skyrmion bags.

\vspace{0.5cm}
\noindent
\textbf{Magnetic properties of Co$_8$Zn$_8$Mn$_4$}

\noindent
The Co$_8$Zn$_8$Mn$_4$ compound crystallizes in the chiral $\beta$-Mn structure, depicted in Fig.~\ref{Fig_1}\textbf{a}.
Like B20-type magnets (e.g.\ FeGe and MnSi), $\beta$-Mn-type Co--Zn--Mn crystals occur in two enantiomorphic forms.
Unlike B20 compounds, whose two enantiomers belong to the same space group ($P2_13$), the two enantiomers of the $\beta$-Mn structure belong to different cubic space groups, $P4_132$ and $P4_332$, corresponding to right- and left-handed crystal structures, respectively.
This distinction originates from the underlying point-group symmetry: the $\beta$-Mn structure belongs to the chiral point group~$O$, whereas B20-type crystals belong to its chiral subgroup~$T$.
Unless stated otherwise, all results presented here correspond to right-handed crystals.

According to Ref.~\cite{TokunagaCoZnMn}, the Curie temperature of Co-Zn-Mn alloys strongly depends on their chemical composition and can range from approximately 150 to 450~K.
For the Co$_8$Zn$_8$Mn$_4$ crystal used in this study, the Curie temperature is $T_\mathrm{c}=350\pm5$~K, and the room-temperature saturation magnetization was $M_\mathrm{s}=240$~kA/m, as determined from measurements on the corresponding bulk crystal presented in Fig.~\ref{Fig_1}\textbf{b}.
As in other chiral magnets, the competition between the Heisenberg exchange interaction and the Dzyaloshinskii-Moriya interaction~(DMI) stabilizes a helical magnetic ground state~\cite{r39}. 
When \textbf{k} vector of helix lies in the plane as in inset of Fig.~\ref{Fig_1}\textbf{b}, the helical domains become visible in Lorentz TEM images.
For the Co$_8$Zn$_8$Mn$_4$, the equilibrium helical period was determined to be $L_{\rm D}=140$~nm.
Taking the above data into account, we identify the material parameters of the system, which we use in our micromagnetic simulations. 
For details, see the Methods section. 
When the \textbf{k} vector is perpendicular to the sample plane, the helical state produces no Lorentz TEM contrast because the projected in-plane magnetization is uniform across the field of view.
This provides a favorable platform for the formation and observation of hopfions at zero magnetic field.

\vspace{0.5cm}
\noindent
\textbf{In situ laser excitation in TEM}

\noindent
\noindent
Our Lorentz TEM is equipped with a femtosecond laser, enabling in situ irradiation of Co$_8$Zn$_8$Mn$_4$, laser-induced nucleation of magnetic textures, and direct observation of the resulting magnetic contrast.
The laser pulses had a wavelength of 520~nm, a duration of 300~fs, and linear polarization.

The Co$_8$Zn$_8$Mn$_4$ samples had lateral dimensions of approximately $5\,\mu{\rm m}\times5\,\mu{\rm m}$, whereas the laser spot diameter was several times larger, ensuring nearly homogeneous irradiation of the entire sample.
Two types of samples were investigated: plates of nearly uniform thickness and wedged plates with thicknesses ranging from approximately 155 to 215~nm. 
Experimentally, hopfions were observed only in sample regions with thicknesses greater than approximately 180~nm.
The sample thickness was determined by electron energy-loss spectroscopy (EELS).
Additional details of the sample preparation, characterization, and TEM instrumentation are provided in the Methods section.
Unless stated otherwise, all measurements presented in this work were performed at zero applied magnetic field and at temperatures at or above room temperature.

The experiment revealed that the threshold laser fluence for generating topological spin textures under single-pulse laser excitation increases with sample thickness. Guided by this trend, uniform excitation fluences of 4.8~cm$^{-2}$ and 7.7~cm$^{-2}$ were selectively applied to the thinner (106 and 140~nm) and the thickest (190 and 215~nm) samples, respectively, in subsequent measurements.

\vspace{0.5cm}
\noindent
\textbf{Observation of room-temperature hopfions}

\noindent
Single-pulse laser excitation can induce the formation of multiple magnetic textures that remain stable after the laser pulse is over.
At elevated temperatures, thermal fluctuations are expected to distort the shape of these textures, as illustrated for a hopfion in Fig.~\ref{Fig_1}\textbf{c}.
Nevertheless, we find that hopfions in Co$_8$Zn$_8$Mn$_4$ remain stable up to approximately $320$~K, with no noticeable reduction in their lifetime.
To quantify their thermal stability, we measured the lifetime of individual hopfions in a 190-nm-thick sample as a function of temperature (Table~\ref{Tab1}).
For each temperature, hopfions were generated by laser excitation and subsequently monitored without further irradiation until they spontaneously collapsed.
At temperatures up to 320~K, no collapse was observed during the 120-minute observation window.
Above this temperature, the lifetime decreases rapidly, reaching only a few seconds at 332~K.

\begin{table}[htbp]
\caption{Lifetime of hopfion at elevated temperatures}
\renewcommand{\arraystretch}{1.4}
\begin{ruledtabular}
\begin{tabular}{lcccc}
\multirow{2}{*}{\textbf{Temperature}} & \multicolumn{4}{c}{\textbf{Measurement}} \\
\cline{2-5} 
 & \textbf{1} & \textbf{2} & \textbf{3} & \textbf{4} \\
\hline 
320 K & $>$120min & $>$120min & $>$120min & -- \\
326 K & 7min05s & 1min46s & 11min13s & $>$50min \\
328 K & 3min33s & 19min02s & 4min12s & 2min28s \\
330 K & 2min38s & 2min23s & 37s & 1min23s \\
332 K & 4s & 24s & 9s & 18s 
\label{Tab1}
\end{tabular}
\end{ruledtabular}
\end{table}

Figure~\ref{Fig_1}\textbf{d} shows selected frames from a 16-hour observation at room temperature of isolated hopfions generated by a single laser pulse in a uniform-thickness sample.
The hopfions exhibit Brownian-like motion characterized by random translational and rotational displacements.
Such stochastic dynamics originate from interactions between the magnetic texture and thermally excited magnons.
Consequently, under room-temperature conditions, the hopfion position remains essentially unchanged on time scales of a few seconds, whereas measurable translational and rotational displacements accumulate only over periods of several minutes or even hours.
Notably, the observation was terminated only because the allocated microscope time had elapsed, while the hopfions remained stable.

Although moderate thermal fluctuations perturb the magnetic textures, they do not alter their topological invariants.
Consequently, we classify the observed magnetic textures using homotopy theory~\cite{Hatcher}.
The topological solitons considered in this work are stabilized within non-collinear magnetic phases, such as helical and conical spin modulations.
In an ideal helical or conical state, the magnetization vectors sweep out a circle on the spin sphere~$\mathbb{S}^2$.
Thermal fluctuations, slight tilts of the modulation axis, and other imperfections broaden this circle into a ring-like region.
Topologically, this region is homeomorphic to a spherical segment and, therefore, to the doubly punctured sphere $\mathbb{S}^2\setminus\{P_1,P_2\}$.
Isolated hopfions, such as the one shown in Fig.~\ref{Fig_1}\textbf{c}, are characterized by the Hopf index~$H$, which belongs to to the third relative homotopy group~\cite{Chen2026hopfions}
\begin{align}
\pi_3(\mathbb{S}^2,\mathbb{S}^2\setminus\{P_1,P_2\})=\mathbb{Z}\quad(\ni\!H).
\label{Z}
\end{align}
The approach used to compute the Hopf index of magnetic textures embedded in a helical background is described in Ref.~\cite{Chen2026hopfions}.
For the experimentally observed hopfion in Fig.~\ref{Fig_1}\textbf{d} and its corresponding digital twin in Fig.~\ref{Fig_1}\textbf{c}, the Hopf index was calculated to be $H=-1$. 

\begin{figure}[!ht]
\centering
\includegraphics[width=1.0\linewidth]{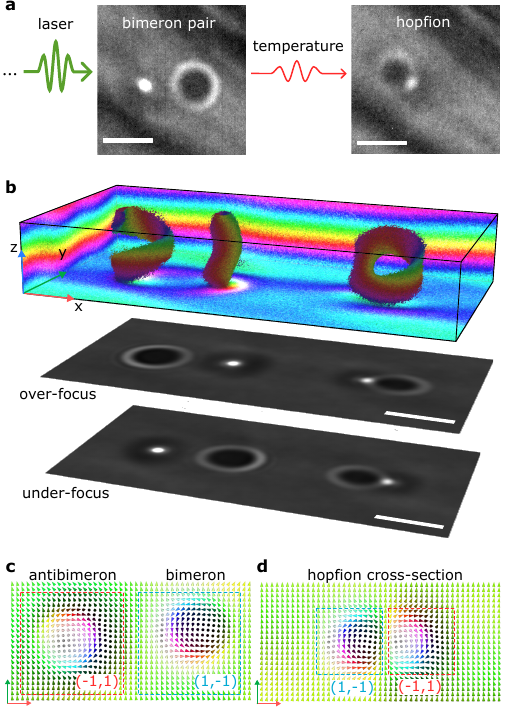}
\caption{\textbf{Bimerons and their fusion into a hopfion.}
    \textbf{a}, Under-focus Lorentz TEM images recorded before and after the fusion of a bimeron and an antibimeron into a hopfion. 
    Before the fusion, the bimeron and antibimeron exchange their relative positions.
    \textbf{b}, Digital twins of a bound pair of oppositely charged bimerons and a hopfion obtained by micromagnetic simulations and shown together for comparison within a single 200~nm-thick slab.
    Thermally distorted isosurfaces for $m_z=-0.6$ and the magnetization shown on the boundaries of the computational box illustrate temperature-induced fluctuations.
    The bottom insets show the corresponding Lorentz TEM images calculated with a defocus of $\pm1\,\mu$m.
    Scale bars in all Lorentz TEM images are 200~nm.
    \textbf{c,d}, Magnetization distribution in the central $xy$ plane of the configurations shown in \textbf{b}, corresponding to the bound bimeron pair (\textbf{c}) and the hopfion (\textbf{d}).
    The pair of indices $(q_\mathrm{t},q_\mathrm{b})$ inside the dashed rectangles denotes the meron topological charges.
}
    \label{fig_bimerons_fusion}
\end{figure}

\vspace{0.5cm}
\noindent
\textbf{Hopfion formation through bimeron fusion}

\noindent
From a technical point of view, direct visualization of the nucleation process following a laser pulse is challenging because the nanosecond timescale is inaccessible to TEM, which typically requires exposure times of milliseconds to seconds.
However, by observing laser-induced bimerons~\cite{Zhu2026bimeron}, we captured their thermally driven fusion into a hopfion, as shown in Fig.~\ref{fig_bimerons_fusion}\textbf{a}.
In our previous study~\cite{Chen2026hopfions}, such a fusion mechanism was proposed as a likely pathway for hopfion formation in chiral magnets.
Such fusion can be understood in terms of bimeron topology.

\begin{figure*}[ht]
    \centering
    \includegraphics[width=\linewidth]{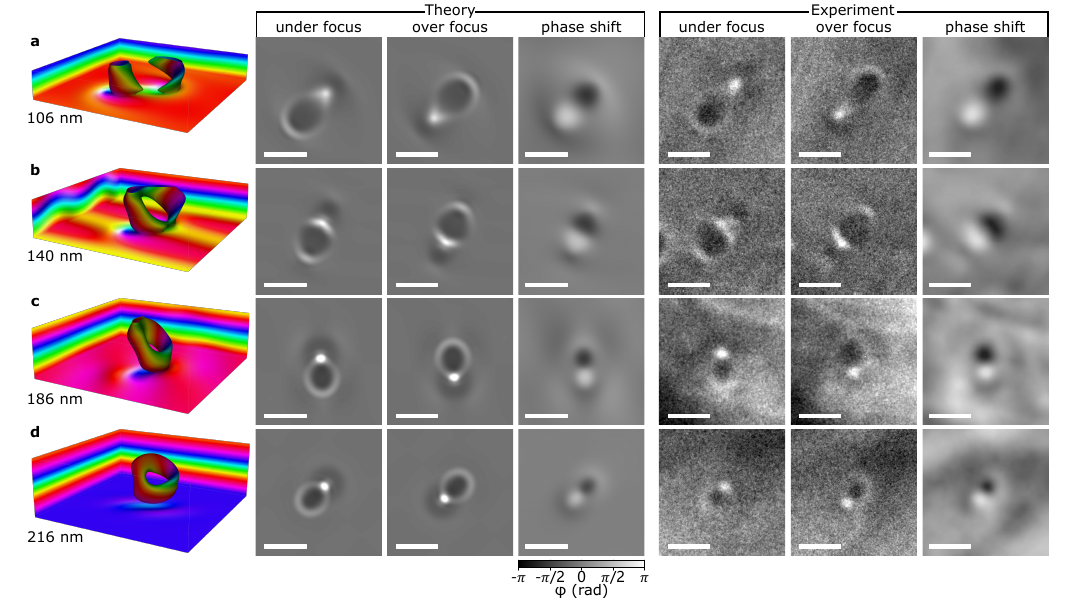}
    \caption{
    \textbf{Statically stable intermediate configurations between bimeron-antibimeron pair and hopfion.}
    \textbf{a}--\textbf{d} display the equilibrium magnetic configurations in samples with thicknesses of 106~nm, 140~nm, 186~nm, and 216~nm, respectively. From left to right, each row presents: the simulated domain visualized via $m_z = -0.5$ isosurfaces (where the side colors represent the perpendicular helix background); the simulated over- and under-focus Lorentz TEM images alongside the electron phase-shift images; and the experimental over- and under-focus Lorentz TEM images coupled with the TIE-reconstructed electron phase-shift images. Scale bars in all TEM images are 200~nm.
    }
    \label{fig_intermediate_forms}
\end{figure*}

A bimeron in a 3D sample consists of a meron--antimeron string localized in two dimensions but extending along the third dimension (through the sample thickness), see Fig.~\ref{fig_bimerons_fusion}\textbf{b}, \textbf{c}.
These bimeron strings were originally described as skyrmion strings embedded in a conical state~\cite{PhysRevLett.115.117201}, and can be characterized by a single topological invariant, the skyrmion topological index~$Q$.
This topological classification was generalized in Ref.~\cite{Chen2026hopfionring} to distinguish skyrmions and bimerons embedded in a helical background.
Unlike the Hopf-index classification introduced in the previous section, the classification provided in Ref.~\cite{Chen2026hopfionring} accounts not only for hopfions localized in all three dimensions but also for meron and skyrmion strings extending through the entire sample thickness.
The corresponding classification is based on a triplet of topological invariants
\begin{align}
(q_\mathrm{t},q_\mathrm{b},h)\in\mathbb{Z}^3.
\label{ZZZ}
\end{align}
Importantly, the group of triplets in Eq.~(\ref{ZZZ}) is defined for a given number of helical pitches, $v$.
The indices $q_\mathrm{t}$ and $q_\mathrm{b}$ denote the meron topological charges~\cite{Rybakov2025Topological} 
and are indicated in Fig.~\ref{fig_bimerons_fusion}, \textbf{c} and {d}.
The third invariant, $h$, is closely related to the Hopf index.
The conventional skyrmion and Hopf invariants, $Q$ and $H$, are recovered from this more general classification defined by Eq.~(\ref{ZZZ}).
For example, when the conical state is continuously transformed into the uniformly magnetized state pointed along the $+z$ direction, the corresponding group homomorphism gives~\cite{Chen2026hopfionring}
\begin{equation}
(q_\mathrm{t},q_\mathrm{b},h)\mapsto
\left(
\underbrace{q_\mathrm{b}}_{Q},
\underbrace{h-vq_\mathrm{b}}_{H}
\right).
\end{equation}

Within the classification defined by Eq.~(\ref{ZZZ}), bimerons carry topological charges $(\pm1,\mp1,-v)$.
The connection to the Hopf index is established by the natural group homomorphism
\begin{equation}
\begin{aligned}
\mathbb{Z}\times\mathbb{Z}\times\mathbb{Z}
&\rightarrow \mathbb{Z},\\
(q_\mathrm{t},q_\mathrm{b},h)
&\mapsto i_\mathrm{t}q_\mathrm{t}+i_\mathrm{b}q_\mathrm{b}+h,
\label{ZZZ_Z}
\end{aligned}
\end{equation}
where coefficients $i_\mathrm{t}$ and $i_\mathrm{b}$ are integers.
This homomorphism establishes the correspondence between the generalized topological classification based on the triplet $(q_\mathrm{t},q_\mathrm{b},h)$ and the Hopf index given by Eq.~(\ref{Z}).
For the experimentally observed fusion of an antibimeron and a bimeron,
\[
(-1,1,-v)+(1,-1,-v)=(0,0,-2v),
\]
which, according to Eq.~(\ref{ZZZ_Z}), yields a hopfion state with Hopf index $H=-2v$.
Thus, the fusion of a bimeron--antibimeron pair generates two elementary hopfions per helical pitch.
In the present experiment, the sample thickness in the region where hopfions are observed is approximately one helical period ($v=1$).
Because the free sample surfaces break the translational periodicity assumed in the above classification, only one of the two hopfions is stabilized, consistent with the experimental observations shown in Fig.~\ref{fig_bimerons_fusion}\textbf{a}.
This interpretation is further supported by earlier experiments on ultrathin FeGe samples (with effective thickness $t\approx\tfrac{1}{2}L_\mathrm{D}$), in which bimeron and antibimeron fusion did not produce hopfions but instead resulted either in a bound bimeron pair or in complete annihilation~\cite{Zheng2022sk-askpairs}.

Translation of a hopfion along the helix axis is intrinsically coupled to its rotation, resulting in screw motion, the general form of a proper motion of 3D Euclidean space~\cite{Kostrikin-Manin}.
Therefore, the orientation of a hopfion is uniquely determined by the local phase of the surrounding helical background.
In a finite sample, this phase is fixed by the sample geometry and magnetic history (e.g., the direction and strength of the previously applied magnetic field).
Hopfions located at the same position along the helix axis therefore have the same orientation.
Since hopfions in relatively thin plates ($t\sim L_\mathrm{D}$) are confined to a narrow region around the sample mid-plane, they all have nearly the same orientation (Fig~\ref{Fig_1}\textbf{d}).
For the hopfion shown in Fig.~\ref{fig_bimerons_fusion}\textbf{b}, the corresponding cross-section through the middle of the sample consists of a $(1,-1)$ bimeron on the left and a $(-1,1)$ bimeron on the right (Fig.~\ref{fig_bimerons_fusion}\textbf{d}).
Therefore, only a bimeron--antibimeron pair with the same mutual orientation can transform into this hopfion.
A bimeron--antibimeron pair with the opposite mutual orientation (Fig.~\ref{fig_bimerons_fusion}\textbf{a,c}) forms another stable configuration and does not merge into a hopfion unless the two textures first exchange their positions, as observed experimentally.
The above scenario applies only to relatively thin plates, where hopfions are confined to the vicinity of the sample mid-plane.
In thicker samples, hopfions can occupy different positions along the helix axis and therefore appear with different orientations.
Conversely, the relative displacement of two hopfions along the helix axis can be estimated directly from their relative orientation.

\vspace{0.5cm}
\noindent
\textbf{Transient bimeron-hopfion textures observed at different thicknesses}

\noindent
Although the details of the fusion process (Fig.~\ref{fig_bimerons_fusion}\textbf{a}) cannot be resolved in TEM directly, the same process can be investigated from a different perspective.
By examining laser-induced textures in samples of different thicknesses, we identified several stable configurations that resemble intermediate stages of the bimeron-to-hopfion transformation.
Such intermediate stages are shown in Fig.~\ref{fig_intermediate_forms}, where each row corresponds to a specific thickness.
The first column shows the corresponding magnetic configurations visualized by the $m_z=-0.5$ isosurface.
The last three columns present the corresponding experimental Lorentz TEM images together with the reconstructed electron phase shift.
The close similarity between the simulated and experimental Lorentz TEM images demonstrates that these micromagnetic configurations serve as effective digital twins of the observed magnetic textures.

\begin{figure*}[ht]
    \centering
    \includegraphics[width=\linewidth]{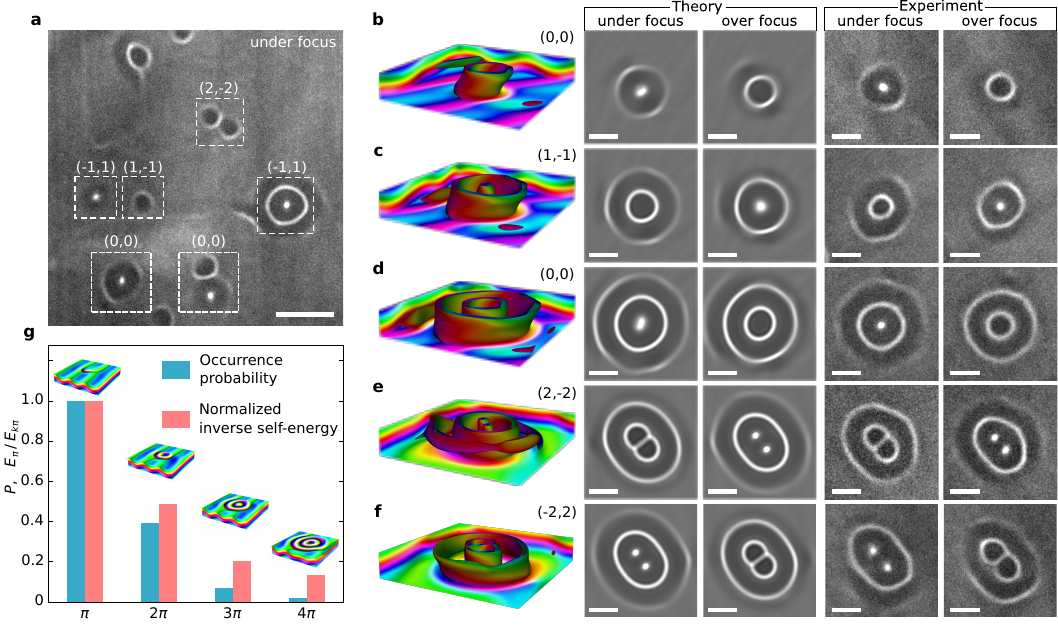}
    \caption{\textbf{Light-induced composite magnetic textures.} 
    \textbf{a} Under-focuse Lorentz TEM image of the 140-nm-thick sample showing a wide diversity of magnetic configurations induced by a single laser pulse of fluence 4.8 mJ/cm$^2$.
    \textbf{b}--\textbf{f} display the equilibrium configurations of various composite magnetic textures obtained in the 140~nm thick sample. The indices in parentheses are the meron topological charge of the corresponding texture, $(q_t,q_b)$. From left to right, each row displays: the magnetization in the simulated domain visualized by ${m_z = -0.5}$ isosurfaces, followed by theoretical and experimental over- and under-focus Lorentz TEM images. All experimental and simulated images share identical 200~nm scale bars.
    \textbf{g}, Experimentally measured occurrence probability of the bimeron and the composite textures shown in \textbf{b--d}, compared with the normalized inverse self-energy, $E_{\pi}/E_{k\pi}$, obtained from micromagnetic simulations.
    }
    \label{fig-4}
\end{figure*}

The 106-nm-thick sample hosts a bound bimeron--antibimeron pair in which the bimeron and antibimeron remain clearly separated (Fig.~\ref{fig_intermediate_forms}\textbf{a}).
Increasing the thickness to 140~nm strengthens the connection between the two bimerons, producing a more compact 3D texture, although a hopfion has not yet formed (Fig.~\ref{fig_intermediate_forms}\textbf{b}).
At a thickness of 186~nm, the first hopfion with Hopf index $H=-1$ emerges, while remaining attached to the sample surfaces (Fig.~\ref{fig_intermediate_forms}\textbf{c}).
With further increasing thickness, the hopfion gradually migrates into the interior of the sample, ultimately becoming largely localized within the bulk (Fig.~\ref{fig_intermediate_forms}\textbf{d}).
Together, these configurations reconstruct the fusion pathway from a bound bimeron--antibimeron pair to a hopfion, with sample thickness replacing time.

Notably, we experimentally observed another configuration of the bimeron--antibimeron pair with a different mutual orientation.
In contrast to the configuration discussed above, when the bimeron and antibimeron exchange positions, they remain separated as the sample thickness increases and do not merge into a hopfion.
Micromagnetic simulations reproduce this behavior and show that the two textures can be observed across the entire thickness range.
These observations demonstrate that only one mutual orientation of the bimeron and antibimeron leads to hopfion formation, consistent with the fusion pathway discussed above.

\vspace{0.5cm}
\noindent
\textbf{Light-induced composite magnetic textures}

\noindent
In addition to hopfions, merons, and their transient states, single-pulse laser excitation generates a wide variety of other textures in a 140-nm-thick sample (Fig.~\ref{fig-4}).
These textures can be broadly grouped into two families resembling $k\pi$-skyrmions~\cite{Bogdanov_99,Zhengtargetskyrmion_1,Zhangtargetskyrmion_2} and skyrmion bags~\cite{Rybakovskyrmionbags,Foster_skyrmionbags,Tangskyrmionbundles,Zhangskyrmionbundles,Liuskyrmionbags,Yangskyrmionbags,Kernskyrmionbags}.
Representative examples of textures resembling $2\pi$-, $3\pi$-, and $4\pi$-skyrmions are shown in Fig.~\ref{fig-4}\textbf{b--d}, while Fig.~\ref{fig-4}\textbf{e,f} shows two representative textures resembling skyrmion bags of opposite topological indices.
These composite textures consist of combinations of merons and antimerons embedded in a perpendicular helical background.
All textures resembling $k\pi$-skyrmions can occur in three configurations characterized by the following values for topological indices $(q_t,q_b)$: $(1,-1)$, $(-1,1)$, and $(0,0)$.
The overall topological index is determined solely by the central magnetic entity, while the surrounding ring-like textures are topologically trivial.

The occurrence probability of magnetic textures resembling $k\pi$-skyrmions was estimated from a statistical analysis of 2000 single-pulse laser excitation experiments performed at room temperature using a laser fluence of $4.8~\mathrm{mJ/cm^2}$.
Following each laser pulse, all magnetic textures within the field of view were classified according to the categories shown in Fig.~4.
Whenever a given configuration was observed, one occurrence was assigned to the corresponding category, irrespective of the presence of other configurations in the same image.
The resulting occurrence probability decreases rapidly with increasing structural complexity and the size of the texture (Fig.~\ref{fig-4}\textbf{g}).
The experimentally measured occurrence probability correlates closely with the inverse relative self-energy, $E_{\pi}/E_{k\pi}$, obtained from micromagnetic calculations (Fig.~\ref{fig-4}\textbf{g}).
These calculations therefore provide a natural explanation for the experimentally observed decrease in the probability of generating increasingly large and energetically unfavorable composite textures.

\vspace{0.5cm}
\noindent
\textbf{Discussion and outlook}

\noindent
We have demonstrated room-temperature magnetic hopfions in Co$_8$Zn$_8$Mn$_4$ that can be generated by single-pulse laser excitation and remain stable in the absence of an applied magnetic field.
Direct observation of the transition from a bimeron pair to a hopfion reveals a formation pathway for 3D topological solitons in chiral magnets.
Our work establishes an experimental platform for investigating the fundamental physics of magnetic hopfions and for testing theoretical proposals for their use in transport, magnonics, and information-processing applications, including Hall effects, racetrack memories, and neuromorphic information processing~\cite{PhysRevLett.123.147203, PhysRevLett.124.127204, PhysRevLett.131.166702, PhysRevResearch.2.013315, ScienceAdvances2023, Waseer2026}.

\section*{Methods}
\noindent\textbf{TEM sample preparation:} Co$_8$Zn$_8$Mn$_4$ thin lamellae were fabricated from bulk crystals using a scanning electron microscopy-focused ion beam (SEM-FIB) dual-beam system (JIB-4700F Multi Beam System) equipped with a gas injection system and a micromanipulator. A lamella with an initial thickness of ${\sim 1}$~$\mu$m was extracted from the bulk crystal using the lift-out method, then transferred onto a copper lift-out grid via the deposited Pt layers. Depending on the experimental requirements, the samples were subsequently thinned to a final thickness of 100–220~nm. A low ion beam current was applied for final polishing to minimize the surface amorphous layer.

\vspace{0.5cm}
\noindent\textbf{Magnetic imaging and femtosecond laser excitation:} Fresnel-defocus Lorentz TEM was employed for magnetic domain analysis using a JEOL JEM-2100 Plus microscope operated at 200~kV. The objective lens current was precisely regulated to apply out-of-plane magnetic fields ranging from $-1.0$~T to $+1.0$~T normal to the specimen plane. All images were acquired at room temperature using a Gatan OneView $4\mathrm{k} \times 4\mathrm{k}$ detector, with the defocus distance maintained at 1~mm unless otherwise specified. The JEM-2100 Plus microscope was retrofitted with a femtosecond laser system (Spirit 1040-4, Spectra-Physics), enabling direct generation of magnetic textures in $\mathrm{Co}_8\mathrm{Zn}_8\mathrm{Mn}_4$ thin plates via pulsed laser excitation. Linearly polarized 520~nm femtosecond pulses ($300$~fs FWHM duration) were focused to a ${\sim 40}$~$\mu$m FWHM spot on the specimen. These pulses were generated by frequency-doubling the fundamental 1040~nm output in a $\beta$-barium borate (BBO) crystal, with pulse energy and repetition rate digitally controlled.

\vspace{0.5cm}
\noindent\textbf{Characterization of samples with different thicknesses.}
The thickness of all TEM lamellae was determined by STEM-EELS using the log-ratio method, with spectra collected across the entire sample region at a spatial step of 100~nm.
The average thicknesses of these regions are approximately 106, 140, and 215~nm, corresponding to the bimeron pair, intermediate, and hopfion configurations discussed in the main text (Fig.~\ref{fig_intermediate_forms}).
A wedge-shaped sample with a thickness gradient was characterized in the same manner. 
The thickness varies continuously from approximately 155 to 215~nm.
Combining observations of fixed-thickness and wedged samples enabled us to identify distinct intermediate stages of the bimeron-to-hopfion transition across different thicknesses.

\vspace{0.5cm}
\noindent\textbf{Hopfion lifetime measurements.}
The sample temperature was stabilized at the desired value before each measurement.
The temperatures reported in Table~\ref{Tab1} correspond to the average sample temperatures, with fluctuations of approximately ${\pm 0.2}$~$^\circ$C during each measurement.
Hopfions were generated by applying femtosecond laser pulses until a well separated from other textures hopfion was observed in Lorentz TEM.
The laser irradiation was then switched off, and the selected hopfion was continuously monitored without further excitation until it spontaneously disappeared.
The elapsed time between the last laser pulse and the spontaneous collapse of the hopfion was defined as its lifetime.
For each temperature, the measurement was repeated up to four times using independently generated hopfions.
At 320~K, none of the observed hopfions collapsed during the 120-minute observation period; therefore, only lower bounds for their lifetime are reported.

\vspace{0.5cm}
\noindent\textbf{Micromagnetic simulations:} 
In this work, we follow the micromagnetic model approach~\cite{MicromagneticsRevisited, zheng2021braids} with the total energy of the system, including four energy terms: the Heisenberg exchange energy, DMI energy, Zeeman energy, and self-energy of the demagnetizing field:
\begin{align}
E=\int\limits_{V_m}\mathrm{d}\mathbf{r}\,\mathcal{A}\sum_{i=x,y,z}&{\lvert \nabla m_{i}\rvert}^2+\mathcal{D}\,\mathbf{m}\cdot(\nabla\times\mathbf{m})-M_{\rm s}\mathbf{m}\cdot\mathbf{B}\nonumber
\\&+\frac{1}{2\mu_{0}}\int\limits_{\mathbb{R}^3}\mathrm{d}\mathbf{r}\sum_{i=x,y,z}{\lvert \nabla A_{d,i}\rvert}^2, 
\end{align}
where $\textbf{m}(\textbf{r})=\textbf{M}(\textbf{r})/M_{\rm s}$ is a unit vector field that defines the direction of the magnetization, $M_{\rm s}=\lvert\textbf{M}(\textbf{r})\rvert$ is the saturation magnetization, $\mathcal{A}$ is the exchange stiffness constant, $\mathcal{D}$ is the constant of isotropic bulk DMI, and $\mu_{0}$ is the vacuum permeability. 
In our simulations, we used the following material parameters for Co$_8$Zn$_8$Mn$_4$: $\mathcal{A} = 5.8$~pJ/m, $\mathcal{D} = 0.52$~mJ/m$^2$, and $M_{\rm s} = 240$~kA/m. 
The value of $M_{\rm s}$ corresponds to room temperature, as shown in Fig.~\ref{Fig_1}\textbf{b}. 
With these parameters, the equilibrium period of helix modulations at zero field is $L_\mathrm{D} = 4\pi \mathcal{A}/\mathcal{D} = 140$ nm, in agreement with the experiment. 
The saturation field of the conical phase~\cite{zheng2021braids} is given by $B_\mathrm{c} = B_\mathrm{D} + \mu_0 M_\mathrm{s} \approx 400$~mT, where $B_\mathrm{D} = \mathcal{D}^2 / (2 M_\mathrm{s} \mathcal{A})$.

The simulation domain was set to $2\,\mu{\rm m} \times 2\,\mu{\rm m}$ in the $xy$-plane, with the thickness varied between $90$~nm and $200$~nm. 
The discretization size in the finite-difference scheme was $4\,{\rm nm} \times 4\,{\rm nm} \times 4\,{\rm nm}$. 
To approximate an extended plate, periodic boundary conditions were applied in the $xy$-plane. 
The FIB-damaged surface layer, $8$~nm thick, was modeled as two cuboid layers at the top and bottom surfaces, where the DMI coupling constant, $\mathcal{D}$, is set to zero~\cite{Zhengbraids,Yangskyrmionbags}.

For the calculation of the self-energy of the $k\pi$-like composite textures shown in Fig.~4\textbf{g}, we used a domain of size $0.8\,\mu{\rm m} \times 0.8\,\mu{\rm m} \times 140\,{\rm nm}$ with periodic boundary conditions in the $xy$ plane.
Although the demagnetizing field was fully accounted for during energy minimization, its contribution to the total energy was excluded from the self-energy analysis; only the exchange and DMI contributions were included.
This avoids an artificial dependence of the calculated self-energy on magnetostatic interactions between periodic images, which become increasingly important for larger magnetic textures within a fixed simulation domain.

All simulations, including theoretical Lorentz TEM images, were performed using the Excalibur code~\cite{excalibur, Zhengbraids}. 
Several solutions were independently reproduced with MuMax3~\cite{MuMax3}. 
Energy minimization was carried out using the conjugate gradient method and other gradient-based methods, as implemented in Excalibur and MuMax3, respectively. 
Consistent with the experimental setup, all simulated Lorentz TEM images were computed for a defocus distance of~${\pm 1}$~mm.

\section*{Data availability}
The data that support the findings of this study are available from the corresponding authors upon reasonable request.

\section*{Acknowledgements}
This work was supported by the National Key R \& D Program of China, Grant Nos.~2024YFA1611303 (S.S.S and Y.Z.), 2024YFA1408701 (J.Q.L.) and 2024YFA1408403 (H.X.Y); the National Natural Science Foundation of China, Grant Nos. 12364018 (Z.A.L.) and U22A6005 (J.Q.L., J.L., H.X.Y. and H.F.T.); the Scientific Instrument Developing Project of the Chinese Academy of Sciences, Grant No. YJKYYQ20200055 (J.Q.L.); the Natural Science Foundation of Guangxi Province, Grant No.~2024GXNSFDA010014 (Z.A.L); N.S.K. acknowledges the European Research Council under the European Union's Horizon 2020 Research and Innovation Programme (Grant No.~856538 - project '3D MAGiC'). F.N.R. and O.E. acknowledge the Swedish Research Council (Grant No. 2023-04899). O.E. acknowledges eSSENCE, STandUPP and the European Research Council (FASTCORR Project No. 854843), WISE – the Wallenberg Initiative Materials Science, funded by the Knut and Alice Wallenberg Foundation and the Knut and Alice Wallenberg Foundation from the Projects No. KAW 2022.0108 and No. KAW 2022.0252. This work was carried out at the Synergetic Extreme Condition User Facility (SECUF, \href{https://cstr.cn/31123.02.SECUF}{https://cstr.cn/31123.02.SECUF}).

\section*{Author contributions}
J.Q.L. supervised the project. K.X.Z., N.S.K. conceived the idea and designed the experiments. K.X.Z., Z.W. and J.X.T. fabricated the Co-Zn-Mn thin plates. K.X.Z., W.L.G., S.S.S., and S.Y.H. performed the TEM experiments. F.N.R. and N.S.K. developed the theory and performed the numerical simulations. K.X.Z., F.N.R., N.S.K. and Z.A.L. performed the data analysis and wrote the manuscript. All authors discussed the results and contributed to the preparation of the manuscript.

\section*{Competing interests}
The authors declare no competing interests.

\bibliographystyle{unsrt}
\bibliography{refs_v1.bib}

\end{document}